\documentclass[aps,nofootinbib,twocolumn,longbibliography]{revtex4-1}

\usepackage{amsmath,amssymb}
\usepackage[bookmarks=false,pdfstartview=FitH]{hyperref} 

\def\be{\begin{equation}}
\def\ee{\end{equation}}
\def\nn{\nonumber}
\def\f{\frac}
\def\tf{\tfrac}
\def\sgn{{\rm sgn}}
\def\intT{{}^{\scriptscriptstyle\rm (i)} T}
\def\intP{{}^{\scriptscriptstyle\rm (i)} P}
\def\phyT{{}^{\scriptscriptstyle\rm (m)} T}
\def\phyP{{}^{\scriptscriptstyle\rm (m)} P}

\begin{document}

\title{Discrete Symmetries in Covariant LQG}

\author{Carlo Rovelli} \email{rovelli@cpt.univ-mrs.fr}
\affiliation{Centre de Physique Th\'eorique, Aix-Marseille Univ, CNRS UMR 7332,
Univ Sud Toulon Var, 13288 Marseille Cedex 9, France}

\author{Edward Wilson-Ewing} \email{wilson-ewing@cpt.univ-mrs.fr}
\affiliation{Centre de Physique Th\'eorique, Aix-Marseille Univ, CNRS UMR 7332,
Univ Sud Toulon Var, 13288 Marseille Cedex 9, France}

\begin{abstract}

We study time-reversal and parity ---on the physical
manifold and in internal space--- in covariant loop gravity. 
We consider a minor modification of the Holst action which
makes it transform coherently under such transformations.
The classical theory is not affected but the quantum theory 
is slightly different.  In particular, the simplicity constraints are slightly 
modified and this restricts orientation flips in a spinfoam to occur 
only across degenerate regions, thus reducing the sources of 
potential divergences.

\end{abstract}


\maketitle

\section{Time Reversal in Tetrad Gravity}
\label{s.intro}

Classically, the physics of gravity is equally well described by the
Einstein-Hilbert action
\be
S_{\scriptscriptstyle E\!H}[g] = \f{1}{2} \int \sqrt{-\det g} \,
R \, d^4 x,
\ee
where the gravitational field is the metric $g$, or by the tetrad action 
\be
S_{\scriptscriptstyle T}[e]  = \int  e^I \wedge e^J \wedge F^{\star}_{IJ},
\ee
where the gravitational field is the tetrad one-form $e$ with components
$e^I = e^I_\mu dx^\mu$ and $F^{IJ}$ are the components of the curvature of
the torsionless spin-connection $\omega = \omega[e]$ determined by the tetrad%
\footnote{We use units where $8 \pi G = 1$. Greek indices are space-time indices while capital
latin indices denoting the 4D internal space are raised and
lowered with the Minkowski metric $\eta_{IJ}$.  The star
indicates the Hodge dual in Minkowski space: $F^{\star}_{IJ}
\equiv {}^\star F_{IJ} := \tf{1}{2} \epsilon_{IJKL} F^{KL}$.
See \cite{Rovelli:2011eq} for the rest of the notation.}.
The relation between the two languages is of course  $g_{\mu\nu} = \eta_{IJ}
e^I_\mu e^J_{\nu}$. These two actions, however, are not equivalent. This
can be seen by performing an internal time-reversal operation
\be
\intT e^0:= -e^0, \qquad \intT e^i:= e^i, \qquad i=1,2,3. \label{t-int}
\ee
Under this transformation, $S_{\scriptscriptstyle E\!H}$ is clearly invariant
as the metric $g = e^I e_I$ is not affected by this transformation, while $S_{\scriptscriptstyle T}$ flips sign,
$S_{\scriptscriptstyle T}[\intT e]\!=\!-S_{\scriptscriptstyle T}[e]$.  The 
difference becomes manifest by writing both actions in tensor notation and in terms of tetrads: 
\begin{eqnarray}
S_{\scriptscriptstyle E\!H}[e] &=& \f{1}{2} \int |\!\det e| \, R[e] \, d^4 x, \\
S_{\scriptscriptstyle T}[e]\   &=& \f{1}{2} \int (\det e) \, R[e] \, d^4 x. 
\end{eqnarray}
They differ by the sign factor
\be
s \equiv \sgn(\det e),  \label{sign}
\ee
where for later convenience we define $\sgn(0) = 0$.

In loop quantum gravity, one utilizes the first order formalism where the
tetrad $e$ and spin connection $\omega$ are treated as independent
variables, and adds to the action the Holst term  
\be
S_{\scriptscriptstyle H}[e, \omega] = \f{1}{\gamma} \int e^I \wedge e^J \wedge F_{IJ},
\ee
which has no effect on the classical equations of motion.  Here we take $\gamma>0$. Thus, the action
usually taken as the starting point for the definition of the quantum theory is   
\begin{eqnarray}  
S[e, \omega]  &=& S_{\scriptscriptstyle T}[e, \omega]
 + S_{\scriptscriptstyle H}[e, \omega] \nn \\
&=&  \int e^I \wedge e^J \wedge \left( F^{\star}_{IJ}
 + \f{1}{\gamma} F_{IJ} \right) \nn \\
&\equiv& \int e^I \wedge e^J \wedge \left(\star
 + \f{1}{\gamma} \right) F_{IJ}.  \label{holst}
\end{eqnarray}
Defining $\intT \omega^{IJ}$ as (note that this is the same transformation as
for $\omega[e]$ in the tetrad action)
\be
\intT \omega^{0i} := -\omega^{0i}, \quad \intT \omega^{ij} := \omega^{ij},
\quad i,j = 1,2,3,
\ee
we observe that the two terms in this action do not transform in the same way
under an internal time reversal:
\be
S[\intT e,\intT\omega]= -S_{\scriptscriptstyle T}[e, \omega]
+ S_{\scriptscriptstyle H}[e, \omega]. 
\ee
That is, $S$ does not transform coherently under $\intT$, in spite of the fact
that this transformation changes only the time orientation of the internal
Minkowski space. Can we replace $S$ with an action that transforms coherently?
This can be done in two different manners: either modifying the first term,
to have it behave as the Einstein-Hilbert action
\be
S'[e, \omega] = \int e^I \wedge e^J \wedge \left(s\star
 + \f{1}{\gamma} \right) F_{IJ};  \label{holst2}
\ee
[recall that $s$ is defined in Eq.\ \eqref{sign}] 
or modifying the Holst term obtaining an action
that changes sign under internal time reversal: 
\be
S''[e, \omega] = \int e^I \wedge e^J \wedge \left(\star
 + \f{s}{\gamma} \right) F_{IJ}.  \label{holst1}
\ee 
In this paper we explore the consequences of both of these corrections upon
quantization.  We build on the recent work of Yasha Neiman 
\cite{Neiman:2011gf} and Jon Engle \cite{Engle:2012yg}, but also on references 
\cite{ Conrady:2008ea,Mikovic:2011bh} where these internal discrete symmetries
have been studied in the context of spinfoams.

Before closing this section, we add a few comments.

\begin{enumerate}

\item {\bf Other definitions of time reversal.}  There also exists a time-reversal 
transformation that acts on the manifold (considered in the spinfoam context 
in \cite{Han:2011as}), defined by 
\be
\qquad \:\:\:\: \phyT e_a^I:= e_a^I, \quad \phyT e_t^I:= -e_t^I,
\quad a=1,2,3, \label{t-phy}
\ee
and the ``total" time-reversal transformation $T = \intT \phyT$.  
$T$ is the time-reversal symmetry mostly considered in the literature.  
$S_{\scriptscriptstyle EH}$, $S_{\scriptscriptstyle T}$ are both even
under $T$, while the Holst term is odd. Note also that $Ts = s$.

\item {\bf Orientation.} Alternatively,  $\intT$ and $\phyT$ can be defined
by leaving the fields untouched and flipping the orientation of the 
internal Minkowski space and the spacetime manifold, respectively. 
A change of the orientation flips the sign of the normalization of the
completely antisymmetric Levi-Civita symbols.  Thus, $\intT$ changes the
sign of $s$ and of the Hodge operator $\star$, while $\phyT$ changes
the sign of $s$ and of the integral of a four-form.   It is easy to check 
that these definitions are equivalent to Eqs.\ \eqref{t-int} and \eqref{t-phy}, 
respectively.  Then $T$ corresponds to reversing the
orientation of the Minkowski space and the manifold simultaneously.

\item {\bf Parity.} We have formulated the issue above in terms of time
reversal, but it is similarly possible to do so in terms of parity.
Define
\be
\qquad \:\:\:\:  \intP e^i\!:= - e^i, \qquad \intP e^0\!:= e^0,
\quad i=1,2,3. \label{p-int}
\ee
Since all actions are invariant under $\intP \intT e = -e$, it is clear
that we have the same structure with internal parity transformations as
we had in terms of $\intT$.  We also have the total parity transformation
defined by $P = \phyP \intP$, where $\phyP$ is defined analogously to
\eqref{t-phy} and \eqref{p-int}.  The Holst term changes by a sign under
$P$, and is invariant under $PT$.

\item {\bf The Ashtekar Electric Field.} In canonical loop gravity one
works in the time gauge and chooses a linear combination of the connection
and its Hodge dual as a canonical variable. The corresponding conjugate
momentum is the Ashtekar electric field $E^{ai}$, but (confusingly) one
finds two different expressions for this field in the literature
\cite{ThiemannBook,Rovelli:2004fk} : 
\begin{eqnarray} 
E^{ai}&=&|\!\det e|\ e^{ai}, \ \ \  {\rm or} \nn \\ 
E^{ai} &=& \epsilon^{abc}\, \epsilon^i{}_{jk}\, e^{j}_{b}e^{k}_{c}.
\end{eqnarray}
The two expressions differ by the sign $s$ and can be derived from $S'$ and $S''$,
respectively.

\end{enumerate}

\section{Modified simplicity constraint}
\label{s.bf}

We now explore the effect of taking $S'$ or $S''$ instead of $S$ as the
starting point for deriving the covariant dynamics of loop quantum
gravity. We begin from the effect on the canonical structure. In this section 
we restrict the analysis to the regions where $s\ne0$: we  analyze the
regions where $s=0$, namely where the determinant of the metric vanishes,
at the end of the section. 

Working on a 3d Cauchy surface $\Sigma$, the momentum
conjugate to the connection $\omega$ is the boundary one-form with values
in the $sl(2,\mathbb{C})$ algebra:
\be
\pi^{IJ} =  \left(s\star  + \f{1}{\gamma} \right)(e^I\wedge e^J) \Big|_\Sigma,
\ee
for $S'$ and 
\be 
\pi^{IJ} =  \left(\star  + \f{s}{\gamma} \right)(e^I\wedge e^J) \Big|_\Sigma,
\ee
for $S''$. The tetrad $e$ maps the normal one-form to the boundary surface
to a (timelike) Minkowski vector $n_I$, which allows us to split this
momentum into its electric $K^I=\pi^{IJ}n_J$ and magnetic components
$L^I=-{}^\star\pi^{IJ}n_J$. Since clearly $n_{I}e^{I}|_{\Sigma}=0$, one of
the two terms vanishes in each component, leaving
\be \label{caso2}
K^{I} = s n_J (e^I\wedge e^J)^{\star}\Big|_\Sigma, \quad
L^{I} =  -\f{1}{\gamma} n_J (e^I\wedge e^{J})^{\star}\Big|_\Sigma
\ee
in the first case, and 
\be \label{caso1}
K^{I} = n_J (e^I\wedge e^J)^{\star}\Big|_\Sigma, \quad
L^{I} =  -\f{s}{\gamma} n_J (e^I\wedge e^J)^{\star}\Big|_\Sigma
\ee
in the second. $K^{I}$ and $L^{I}$ are normal to $n_{I}$ and live therefore
in a 3d space (oriented by the $n$ and the orientation of $\cal M$). We use
the notation $\vec K=\{K^{i}, \ i=1,2,3\}$ to indicate them.  
For $S'$, equation \eqref{caso2} implies 
\be \label{simpl-eq2}
\fbox{\parbox{3cm}{$\ \ \ \,  \vec K+ s \gamma \vec L= 0$}}
\ee
while for $S''$, equation \eqref{caso1} 
gives 
\be \label{simpl-eq1}
\fbox{\parbox{3cm}{$\ \ \ \, s \vec K+ \gamma \vec L= 0$}}
\ee
which is equivalent to \eqref{simpl-eq2} since $s=\pm1$.
This is the modified linear simplicity constraint for the
actions $S'$ and $S''$.%
\footnote{The conventional action $S$ gives the linear
simplicity constraint $\vec K + \gamma \vec L= 0$ instead
\cite{Engle:2007wy,Freidel:2007py}. But note that,
consistently with what we find here, a negative sign
is obtained in \cite{Neiman:2011gf}.}

$\vec K$ and $\vec L$ are two-forms on the oriented 3d space $\Sigma$, that
is, for instance $K^i=K^i_{ab}dx^a\wedge dx^b$. Therefore they define $3\times3$
matrices, like $K^{ci}:=K^i_{ab}\epsilon^{abc}$, whose determinant we
indicate, respectively, as $\det K$ and $\det L$.  Let
$s_{\!\scriptscriptstyle K}:=\sgn [\det K];$ and
$s_{\!\scriptscriptstyle L}:=\sgn [\det L]$. Since in these coordinates
we have $e^0|_\Sigma=0$, one can check easily that 
\be \label{due}
s = s_{\!\scriptscriptstyle K}, \qquad s_{\!\scriptscriptstyle L} = 1
\ee
in the $S'$ case%
\footnote{In this case, $s_{\!\scriptscriptstyle L} = 1$ implies that
$\vec L$ is a pseudovector with respect to $\intT$ and $\intP$, since
it does not change sign under parity and time reversal, while $\vec K$
is a proper vector as its determinant can be positive or negative.  In
the quantum theory $\vec L$ generates rotations and $\vec K$ boosts,
thus $S'$ appears to better respect the expected transformation properties
of $\vec L$ and $\vec K$.},
while in $S''$
\be \label{uno}
s = s_{\!\scriptscriptstyle L}, \qquad s_{\!\scriptscriptstyle K} = 1.
\ee

So far we have only considered the nondegenerate case.  The degenerate case
occurs when $\vec K = \vec L = 0$ and therefore $s_K = s_L = 0$ as well.
For the degenerate sector, the simplicity constraints have the same form
for the two actions $S'$ and $S''$ and are that both $\vec K$ and $\vec L$
must vanish.

\section{Discretization}

As a step towards the quantum theory, consider the discretization of the theory. 
Introduce an (oriented) triangulation of space-time and integrate the
two-forms $\pi$ over two-cells $f$ in the triangulation. This associates the variable
\be 
\pi^{IJ}_f = \int_f \pi^{IJ}
\ee
to each $f$. Consider one such face $f$ sitting on the boundary of the manifold. 
With respect to the frame defined by $n_I$, determined by the normal to the boundary, 
this momentum splits into its electric and magnetic components  $\vec K_f$ and $\vec L_f$. 
Consider a three-cell in $\Sigma$ and let $f_i, \ i=1,2,3$ be three of
its four faces, ordered according to the orientation of the manifold. 
Define
\be 
\det L := \vec L_{n_1} \cdot \vec L_{n_2} \times \vec L_{n_3},
\ee
which is the discrete analog of the determinant of $\vec L$ in the continuum%
\footnote{Note that it is possible to choose any three of the four edges meeting
at the edge (so long as the relative orientation is taken into account) due to
the closure constraint on $\vec L$ (which is equivalent to the Gauss constraint
in the quantum theory).  There is no closure constraint on $\vec K$ in the
quantum theory without the simplicity constraint and therefore defining
$\det K$ in the discrete theory is not useful for spin foam models.},
and the sign
\be \label{def-sksl}
s_{\!\scriptscriptstyle L}:=\sgn  [\det L].
\ee
We see that $\vec L$ and $\vec K$ live on faces in the discretized theory
while $s_L$ and $s_K$ are associated to tetrahedra.

With these definitions, it is possible to rewrite the simplicity
constraints in the discrete theory for the nondegenerate case.

For $S'$, we must have $s_{\!\scriptscriptstyle L} = 1$.  Also, the constraint
\eqref{simpl-eq2} becomes
\be \label{dis-simp2}
K_f \pm \gamma L_f = 0,
\ee
as both values of $s_{\!\scriptscriptstyle K} = \pm 1$ are allowed.  The
important point is that, since there is one $s_K$ per tetrahedron, one of
$\pm 1$ must be chosen for all of the four faces that compose each tetrahedon.

For $S''$, we have $s_{\!\scriptscriptstyle L} = \pm 1$ in the nondegenerate
sector, and the constraint \eqref{simpl-eq1} is
\be \label{dis-simp1}
K_f + s_{\!\scriptscriptstyle L} \gamma L_f = 0,
\ee
which implies that $s_{\!\scriptscriptstyle K} = 1$.

Now let us consider the degenerate cases.  A degenerate tetrahedron is one
where $s_K = s_L = 0$, while a degenerate face is one where $K_f = L_f = 0$.
Note that a tetrahedron can be degenerate without its faces being degenerate
(and vice versa) and therefore \emph{a degenerate tetrahedron cannot constrain
its faces}.

Finally, the oriented square volume $V^2$ of a three-cell is
determined by \cite{Rovelli:2011eq}
\be 
V^2 = \f{2}{9} \gamma^{3} \, \vec L_{n_1} \cdot \vec L_{n_2}
\times \vec L_{n_3},
\ee
which gives the important relation
\be \label{def-sl}
s_{\!\scriptscriptstyle L} = \sgn(V^2).
\ee

\section{Quantum Theory}
\label{s.quant}

Let us now study the effect of using the modified simplicity condition on
the quantum theory.  We refer the readers to \cite{Perez:2004hj,perez,
Rovelli:2011eq,Engle:2007wy} for the general construction. 

In the quantum theory, $\pi^{IJ}_f$ is promoted to a quantum operator
which is identified as the generator of $SL(2, \mathbb{C})$ over a suitable
space formed by $SL(2, \mathbb{C})$ unitary representations. $\vec K_f$ and
$\vec L_f$ are then the generators of boosts and rotations respectively. The
unitary representations of $SL(2, \mathbb{C})$ are labelled by the two
quantum numbers $\rho$ and $k$, where $\rho \in \mathbb{R}^{+}$ and
$2k \in \mathbb{Z}$. A discrete basis in the  $(\rho,k)$ representation
is obtained by diagonalizing the total angular momentum $|\vec L|^2$ of the
rotation subgroup of  $SL(2, \mathbb{C})$ and its $L_{3}$ component. The
basis vectors are then denoted by $|\rho, k; j, m \rangle$, where $j$ is a
half-integer greater or equal to $|k|$ while $m$ is a half-integer in the
interval of $[-j, j]$.  The Casimirs of $SL(2, \mathbb{C})$ are
$C_{1}=\vec L \cdot \vec K$ and $C_{2}=|\vec L|^2 -|\vec K|^2$ and take
the values $C_{1}=\rho k$ and $C_{2}= k^{2}-\rho^{2}$ in the $(\rho, k)$
representation. If the quantum operators $K^i_f$ and  $L^i_f$ are defined
on the representation $(\rho_{f},k_{f})$ and satisfy the modified
simplicity constraint \eqref{dis-simp2} or \eqref{dis-simp1}, the states
in the quantum theory must therefore satisfy the relations (see
\cite{Engle:2007wy} for the details of this procedure)
\be \label{qnum1}
\rho_{f} = \gamma j_{f}; \qquad k_{f} = s \, j_{f},
\ee
where $s$ is a sign coming from \eqref{dis-simp2} or \eqref{dis-simp1}.
As $j > 0$, this relation determines the
sign of the quantum number $k$, which in the literature was usually taken
to be positive (although not in \cite{Neiman:2011gf}). Therefore the key effect
of the introduction of the sign $s$ is that the quantum theory now includes
both positive and negative $k$ representations. 

Thus, given $j_f$ and $s$, it is possible to determine $\rho_f$ and $k_f$.
As one can easily check from \eqref{dis-simp2} and \eqref{dis-simp1},
it is necessary to know $s_L$ in order to implement the simplicity constraints
and therefore, one must calculate $s_L$ for each edge.  In order to do this,
we first diagonalize the state with respect to the operator corresponding to
$V^2$ for each edge, which is equivalent to diagonalizing the states with
respect to the $s_L$ operator given by \eqref{def-sl}.  This determines $s_L$
for every edge and in the next section, we show how to use this in order
to implement the simplicity constraints in the vertex amplitude.

\section{Amplitude}
\label{s.amp}

Let us now see the effect of the above on the amplitude that defines the
quantum theory \cite{Rovelli:2011eq}.  We start by recalling the usual form
of the covariant loop quantum gravity amplitude \cite{Engle:2007uq,Livine:2007ya,
Livine:2007vk, Engle:2007qf, Freidel:2007py, Engle:2007wy, Kaminski:2009fm}.
Among the numerous equivalent manners of writing this amplitude, we choose
the ``Polish" one \cite{Bahr:2010bs}:
Let $\Delta$ be a two-complex with faces $f$, edges $e$ and vertices $v$. For
simplicity we assume here that $\Delta$ is the dual of a 4d triangulation and
without boundaries. The amplitude associated to this triangulation is 
\be \label{amplitude}
A_{{}_\Delta} = \sum_{j_f}\ \mu(j_f) \ \rm{Tr}_{{}_\Delta} \prod_e  P_e.
\ee
Here the half-integer $j_f$ is the assignment of a spin to each face,
$\mu(j_f)=\prod_f(2j_f+1)$ is a measure factor and the operators $P_e$ are
defined on the space 
\be \label{He}
H_e = \otimes_{f\in e} H_f
\ee
where $H_f$ is the Hilbert space carrying the $SL(2,\mathbb{C})$
representation $(\rho,k)=(\gamma j_f, j_f)$.   The trace is obtained
by tracing over all Hilbert spaces $H_f$ at couples of edges sharing
the same face at the vertices. The model is then defined by 
\be \label{ghg}
P_e = P_g P_h P_g,
\ee
where $P_g$ is the projection on the $SL(2,\mathbb{C})$ invariant subspace
of $H_e$ (the intertwiner space), and $P_h=(\otimes_{f\in e} P^f_h)$ is
the projection on the $SU(2)$ invariant substance of $H_f$ with $SU(2)$
spin $j_f$. This defines covariant loop quantum gravity. This is the
amplitude that has been shown to be related to the general relativity
action in the large distance limit \cite{Barrett:2009mw,Conrady:2008ea,
Ding:2011hp,Han:2011re}.

Let us now define the variant of the theory that takes the orientation into
account.  The first step is to introduce the projectors $P(s_L = 0, \pm1)$.
The projector $P(s_L = 0)$ annihilates all states with a nonzero volume,
while $P(s_L = 1)$ and $P(s_L = -1)$ project onto the subspaces where
the oriented square volume is positive and negative, respectively.

The next step is to determine the relation between $s_K$ and $s_L$ in
the $SU(2)$ invariant subspace of an $SL(2, \mathbb{C})$ representation.
From the definitions \eqref{def-sksl} and by looking at the action of the
operators $K_i$ and $L_i$ in $SL(2, \mathbb{C})$ given for instance in
\cite{perez}, it is easy to derive that
\be \label{rel-signs}
s_K = \sgn \big( k_{f_1} k_{f_2} k_{f_3} \big) \ s_L,
\ee
where the signs of $k_{f_1}, k_{f_2}, k_{f_3}$ and $k_{f_4}$ are all
the same, as can be seen from the discretized simplicity constraints
\eqref{dis-simp2} and \eqref{dis-simp1}. An important consequence of
this relation is that for a state where $s_L= 0$,  the relation
$s_K= 0$ also necessarily holds in the $SU(2)$ invariant subspace.

\subsection{The Amplitude for $S'$}

To define the quantum theory for the action $S'$ we have to change the
above definition in order to implement two modifications: (i) $k_f$ should
be allowed positive as well as negative, and (ii) $s_L$, which is equal to
the sign of $V^2$, must be positive.  These are easily implanted by 
defining the amplitude%
\footnote{We can restrict the sum to be over nonzero $k_f$.  Even though
degenerate faces $k_f = j_f = 0$ are allowed by the simplicity constraints,
we know from canonical loop quantum gravity that links with $j = 0$ can
be erased from the spin-network.  The same will be done for the amplitude
of the action $S''$ as well.}
%
\be \label{amplitude-s''}
A_{{}_\Delta}' = \sum_{k_f}\  \mu(j_f)
\ \rm{Tr}_{{}_\Delta} \prod_e  P_e',
\ee
where $j_f = |k_f|$, the operators $P_e'$ are defined on the Hilbert space
\be
H_e' = \otimes_{f\in e} H_f'
\ee
where $H_f'$ is the Hilbert space carrying the $SL(2, \mathbb{C})$
representation $(\rho,k)=(\gamma |k_f|, k_f)$. 
The $P_e'$ operator is defined by
\be 
P_e' = P_g P_h P_s' P_h P_g,
\ee
where $P_s'$ is the additional projector defined as follows:
\begin{align} 
P_s' =& \, P(s_L = 1) \times  \!\!\! \prod_{f_1, f_2 \in e} \!\!\!
\delta \Big( \sgn(k_{f_1}), \sgn(k_{f_2}) \Big) \nn \\
& + P(s_L = 0),
\end{align}
where the Kronecker delta imposes the signs of all of the $k_f$ meeting at
a nondegenerate edge to agree (there is no such constraint for faces meeting
at a degenerate edge).

Since $P_h$ projects on $j_f$, we have immediately that $k_j = s j_f$, where
$s \equiv \sgn(k_f)$ which is the same no matter which face is chosen due to the
Kronecker delta.  It is easy to see from Eq.\ \eqref{rel-signs} that $s_K$
can be positive or negative, as wanted.

Notice that on the one hand the states in the sum have doubled because $k_f$ can take
both signs, but on the other hand they are halved as all states with $s_L = -1$
are killed.

Therefore, for $S'$ it will be necessary to work with Hilbert spaces
carrying the representations $(\rho_f = \gamma j_f, k_f = j_f)$ and
$(\rho_f = \gamma j_f, k_f = -j_f)$.
Note that only the first is considered in the usual EPRL model
\cite{Engle:2007wy}, though in that case there is no projector $P_s'$.
(We shall see that for $S''$ these two representations will again be
needed although the extra projector $P_s''$ is different.)

We close with a brief discussion of the gluing conditions.  By looking at the
vertex amplitude and in particular the form of the projector $P_s'$, it is
easy to see that it is impossible to connect two nondegenerate edges with opposite values of
$s_K$.  However, it is possible to connect degenerate edges with any other
type of edge.  Therefore, regions with an opposite sign of $s_K$ can only
be connected by passing through a ``boundary region'' composed of degenerate
edges.

\subsection{The Amplitude for $S''$}

In this case, we have to change the vertex amplitude in order to implement
the following two modifications: (i) $k_f$ should be allowed positive as
well as negative, and (ii) $s_K$ must be positive.  This can be obtained
by defining
\be \label{amplitude-s'}
A_{{}_\Delta}'' = \sum_{k_f}\  \mu(j_f)  \ \rm{Tr}_{{}_\Delta} \prod_e  P_e''.
\ee
where the operators $P_e''$ are defined on the same Hilbert space as $P_e'$
and the $P_e''$ operator is 
\be 
P_e'' = P_g P_h P_s'' P_h P_g.
\ee
where the new projector $P_s''$ is defined by
\begin{align} 
P_s'' =& P(s_L = 0)
+ P(s_L = 1) \times \prod_{f \in e} \delta \Big( k_f, j_f \Big) \nn \\
& + P(s_L = -1) \times \prod_{f \in e} \delta \Big( k_f, -j_f \Big).
\end{align}
Now there is no restriction regarding the sign of the oriented volume squared
operator, but the simplicity constraint \eqref{dis-simp1} must be imposed.
Due to the relation \eqref{rel-signs}, it follows that $s_K \neq -1$
follows automatically.

Once again, it is easy to see that two regions where the $s_L$ have opposite
signs cannot be glued together directly.  Instead, it is necessary to pass through
a degenerate edge in order to travel from a region with $s_L = 1$ to another
where $s_L = -1$.

Thus, just as for $S'$, there must be a ``bridge'' of degenerate edges between
regions with opposite signs of $s$.  In particular, in a connected, nondegenerate
region, we must have constant $s$ everywhere.  This is not particularly surprising
as it is very similar to what we find in the classical, continuous theory: the
relative orientation between the physical manifold and the internal space can
only change at singularities.

\subsection{Comparison to Similar Results}

In \cite{Neiman:2011gf}, the simplicity constraint was modified from
$\vec K + \gamma \vec L = 0$ to $\vec K - \gamma \vec L = 0$.  This
corresponds to the case here when $s = -1$. In the model
proposed in \cite{Neiman:2011gf}, the value of $s$ can still flip from
one cell to the next without restriction and also the simplicity constraint
is not affected by the value of $s$.  On the other hand, in the models
presented here, the value of $s$ can only change across a degenerate
region and then this change plays a role in the simplicity constraints
\eqref{simpl-eq1}.

Reference \cite{Engle:2012yg} suggests a modification of the quantum theory 
very similar to \eqref{amplitude-s''}.  However, the two differences are:
(i) the additional projector introduced in \cite{Engle:2012yg} (in the
Euclidean setting) is in fact $P(s_L = 1)$ as it only allows states where
$V^2_e > 0$ (note that $V^2_e = 0$ is not allowed, another difference
with the prescription we give here) and (ii) there is no sum over both
signs for $k_f$.  Thus the prescription given in \cite{Engle:2012yg} is
different as it imposes $s_K = 1$ (in addition to $s_L = 1$), while
$s_K = \pm 1$ are both allowed configurations for the amplitude coming
from $S'$ presented in this paper.\footnote%
{The problem raised by the sign of $s$ is related to several 
sign issues that have been discussed in the quantum gravity literature. See 
for instance the analysis of causality in spinfoams in \cite{OL1}; the restriction 
to positive physical-time energy in the reconstruction of the spinfoam formalism 
from loop cosmology \cite{Ashtekar:2010ve};  the need to select a phase 
picking up one component of the amplitude in reconstructing semiclassical
transition amplitudes \cite{Rovelli:2005yj,Bianchi:2006uf,Bianchi:2010zs}; 
the interpretation of the early versions of the bounce loop cosmology \cite{BojowaldLR};
the analysis of parity in the Bianchi models \cite{Ashtekar:2009um}; 
and in certain subtle and controversial points of the canonical 
quantization \cite{Giesel:2005bk}; and the effect of
orientation flip in gluing, for spinfoam amplitudes \cite{Han:2011re}.
 In fact, uncertainties about the physical 
interpretation of this sign factor have been present since the very early 
calculations in loop gravity \cite{Iwasaki:1992qy}.}

\section{Analysis}

The first key consequence of the alternate definitions of the vertex amplitude
given above is that in every connected nondegenerate region, i.e., where
$V^2_e \neq 0$, the sign of $s$ remains constant.  As is clear from
Sec.\ \ref{s.intro}, $s$ indicates the relative orientation between the
physical manifold $\mathcal{M}$ and the auxiliary Minkowski space $M$ and
thus this result indicates that the relative orientation cannot flip
without going through a degenerate region in the triangulation.

In the asymptotic analysis in \cite{Barrett:2009mw}, one finds a sum of
two terms in the semiclassical limit of the amplitude \eqref{amplitude},
\be \label{asym}
\lim_{\rm semi-class.} A_{{}_\Delta} \sim e^{i S_R} + e^{-i S_R},
\ee
where $S_R$ is the Regge action.  These two terms correspond to the two possible
relative orientations between $\mathcal{M}$ and $M$.  In the EPRL model, one of
these two terms appears for each edge, depending upon the relative orientation
chosen at that particular edge.  Neighbouring edges do not need to be
glued consistently and therefore a mixing occurs between the two terms in the
semi-classical limit. This mixing is directly responsible for some of the divergences
in the semiclassical limit \cite{ta}.

The vertex amplitude studied here behaves differently.  In the case $S'$,
one might hope that the restriction to the positive eigenspaces of
$V^2$ selects only one of the two sectors in the saddle point approximation
of the vertex  \cite{Barrett:2009mw, Conrady:2008ea, Ding:2011hp, Han:2011vn, Han:2011re},
leading to just one critical point instead of two,
as the theory $S'$ is essentially the Einstein Hilbert one where the sign
of the action of a time reversed configuration does not flip.  However, a
new critical point might be picked up corresponding to $s_K = -1$,
a configuration that was not included in previous studies.  This critical
point would give the same contribution as the surviving critical point
corresponding to $s_L = s_K = 1$ and then the asymptotics would have the
form
\be \label{asym-s'}
\lim_{\rm semi-class.} A_{{}_\Delta}' \sim 2 \, e^{-i S_R}.
\ee
If so, the action $S'$ would realize the objective sought for in
\cite{Engle:2012yg} (in fact, in a very similar manner), but in a
context in which both orientations exist in the theory.

Considering the properties of $S''$ under the discrete transformations,
we expect both terms to appear (as in the EPRL model)
\be \label{asym-s''}
\lim_{\rm semi-class.} A_{{}_\Delta}'' \sim e^{i S_R} + e^{-i S_R},
\ee
but there is an important
difference between the vertex amplitudes as now each of the terms in the
asymptotic analysis corresponds to a connected, nondegenerate region,
rather than the orientation of each cell.  In other words, the weight
associated to connected regions with non-degenerate configurations should
be the cosine of the total Regge action, and not the product of the
individual cosines of the Regge action of each cell.  This is because
connected edges \emph{must} have the same relative orientation between
$\mathcal{M}$ and $M$, unless they are separated by a degenerate region.

Therefore we still have both orientations playing a role (both in $S'$ and
$S''$), but not cell by cell.  Instead, this would occur patch by patch,
on the basis of connected non-degenerate regions (see \cite{Conrady:2008ea}).

\section{Discussion}
\label{s.dis}

Based on the behaviour of the Holst action under internal time reversal and
parity transformations, we have considered two distinct modifications to the
Holst action which lead to the modified actions $S'$ and $S''$.  In both
cases, the simplicity constraints are slightly changed and the
spinfoam quantization of the actions is a little different from the conventional 
one defined for instance in \cite{Rovelli:2011eq}. These modifications might 
reduce one of the sources of divergences in the semi-classical limit, 
as the relative orientation can flip only across degenerate
regions, thus removing some of the problematic mixing terms \cite{ta}.

The alternative between the actions $S'$ and  $S''$ reflects the alternative between 
the Einstein-Hilbert action $S_{\scriptscriptstyle E\!H}$ and the tetrad action
$S_{\scriptscriptstyle T}$.  In the classical theory the choice does not matter,
but the two actions appear to lead to inequivalent quantum theories.  In a Feynman
sum-over-histories approach, summing over tetrads with both signs of the
determinant in $S_{\scriptscriptstyle T}$ is like considering each metric
spacetime in $S_{\scriptscriptstyle E\!H}$ twice: once future-oriented and
once past-oriented, and weight the two with two opposite signs of the action.
The cosine rather than the exponential that appears in the Ponzano-Regge
asymptotics can be interpreted as having this origin.  In the theory defined by  $S'$
both the future- and past-oriented configurations are summed over, just as
in $S''$, but they are weighted with the same sign.  Is there a reason to
prefer one action rather than the other?

One argument in favour of   $S_{\scriptscriptstyle EH}$ and $S'$ is the consideration that 
under internal time-reversal and parity transformations, the generators of boosts 
transform as proper internal vectors ($\vec K \to - \vec K$) while the generators
of rotations transform as pseudo internal vectors ($\vec L \to \vec L$)
in $S'$, as one would expect on geometric grounds.  The situation is reversed in $S''$ 
where $\vec K$ transforms as a pseudo internal vector and $\vec L$ as a proper internal vector.

One argument by analogy in favour of $S_{\scriptscriptstyle T}$ and $S''$, on the other hand, is the
fact that in non-relativistic physics the action of a trajectory moving backward
in time  and that of the same trajectory going forward have opposite signs.
The action for a process is $S = E \Delta T$, and if $\Delta T$
changes sign, so does $S$. This property is lost in $S_{\scriptscriptstyle E\!H}$
because of general covariance, which implies that there is no way of distinguishing
a forward moving spacetime from backward moving one.  But it is present in
$S_{\scriptscriptstyle T}$ and $S''$ as they depend on the sign of $s$.

We close with a comment on the interpretation of regions
with opposite $s$. In Feynman's picture one obtains 
quantum amplitudes summing over the particle's paths in space.
The idea that in this context particles running backward in time 
represent antiparticles forms the intuitive basis of the St\"uckelberg-Feynman 
form of positron theory \cite{Stuckelberg,Feynman}.
According to a beautiful argument given by Feynman in \cite{FeynmanAntiP}, 
special relativity requires such particles running back in time to exist, if the energy
must be positive.  This is because positive energy propagation spills
necessarily outside the light cone. But a propagation of this kind is spacelike 
and therefore can be reinterpreted as backward in time in a different Lorentz frame. 
Therefore there must exist propagation backward in time in the theory and this
represents a (forward propagating) antiparticle. Thus, according to Feynman, 
the existence of antiparticles follows directly from quantum mechanics 
and special relativity.  Can an analogous argument be formulated in quantum gravity?

Consider a gas of
particles in space-time used to define a physical comoving coordinate system.
These define a time function with respect to which the gravitational field can
be seen as evolving. In the quantum theory, however, the gravitational field can
fluctuate off-shell so that the trajectories are somewhere space-like. But then
there is a coordinatization of space-time with respect to which the particles run
backward in time. In turn, the metric in this coordinatization runs backwards in
time with respect to the time defined by the physical reference field. In other words,
we are again in the situation where a solution running backward in time must be
included in the path integral. These are only speculative remarks, but
they suggest that the contribution of the tetrad fields with negative
determinant ---negative internal time--- should perhaps not be dismissed 
lightly a priori.  

Can this intuition be 
relevant for the dynamics of spacetime itself and shed some light on the physical 
interpretation of a region with a flipped internal time direction? Can a region
with the opposite internal time direction be thought of as a spacetime running backward
in time, or an ``anti-spacetime"?

\acknowledgments

We thank Wolfgang Wieland, Simone Speziale,
Alejandro Perez, Abhay Ashtekar and Leonard Cottrell for helpful discussions.
This work was supported by Le Fonds qu\'eb\'ecois de la recherche
sur la nature et les technologies.


\end{document}